\documentclass[journal]{IEEEtran}
\usepackage[colorlinks,linkcolor=blue,anchorcolor=blue,citecolor=blue]{hyperref}
\usepackage{cite}
\usepackage{url}
\usepackage[caption=false]{subfig}
\usepackage{float}
\usepackage{makecell}

\usepackage{amsmath,amssymb,amsfonts}
\usepackage{hyperref}
\allowdisplaybreaks[4]
\usepackage{graphicx}
\usepackage{enumerate}
\usepackage{array}
\usepackage{multirow}
\usepackage{color}
\usepackage{colortbl}
\usepackage[ruled,vlined,linesnumbered]{algorithm2e}
\usepackage{tikz}
\begin{document}

\title{Real-Time Neural Distributed Energy Resources Dispatch with Feasibility Guarantees}

\author{Jie Zhu,~\IEEEmembership{Graduate Student Member,~IEEE,} Yinliang Xu,~\IEEEmembership{Senior Member,~IEEE,} and Hongbin Sun, \IEEEmembership{Fellow,~IEEE}
        % <-this % stops a space
\thanks{
This work has been submitted to the IEEE for possible publication. Copyright may be transferred without notice.

Jie Zhu and Yinliang Xu are with the Tsinghua Shenzhen International Graduate School, Tsinghua University, Shenzhen 518055, China (Corresponding author e-mail: xu.yinliang@sz.tsinghua.edu.cn).

Hongbin Sun is with the Department of Electrical Engineering, Tsinghua University, Beijing 100084, China.}
}
% The paper headers
\markboth{Submitted}%
{Shell \MakeLowercase{\textit{et al.}}: A Sample Article Using IEEEtran.cls for IEEE Journals}

\IEEEpubid{}
% Remember, if you use this you must call \IEEEpubidadjcol in the second
% column for its text to clear the IEEEpubid mark.

\maketitle

\begin{abstract}
The growing penetration of renewable energy necessitates high-frequency real-time scheduling. While neural network-based surrogates enable computationally efficient scheduling, strictly enforcing nonconvex power flow constraints without external solvers remains a fundamental challenge. To bridge this gap, this letter proposes a solver-free neural dispatch framework with rigorous feasibility guarantees. A convex inner approximation of the DistFlow model is first derived via the convex envelope theorem. Building upon this approximation, a robust optimization-based affine policy is formulated to yield a theoretically certified interior-point mapping rule, which is then embedded within a bisection-based projection scheme to efficiently recover feasibility for infeasible NN outputs without any external solver. Experimental results demonstrate that the proposed method restores feasibility on the order of $10^{-3}$ s while maintaining near-optimal performance.
\end{abstract}


\section{Introduction}
\IEEEPARstart{T}{he} growing penetration of renewable energy introduces rapid generation fluctuations, necessitating high-frequency real-time scheduling to ensure grid reliability and operational efficiency. Neural network (NN)-based surrogates offer a promising approach for efficient real-time dispatch. However, a fundamental challenge lies in ensuring that the NN outputs strictly satisfy the underlying physical constraints. 

A common approach is to incorporate violations as penalty terms into the loss function \cite{10540179}. While simple to implement, this strategy cannot strictly enforce feasibility, often yielding infeasible solutions. To enforce hard feasibility guarantees, a direct approach is to project infeasible solutions onto the feasible set by solving a projection optimization problem~\cite{10411984}. However, this relies on an external optimization solver, undermining computational efficiency. While feasibility without solver has been achieved for linear constraints~\cite{10256159,constante2025enforcing}, the inherent nonconvexity of power flow constraints renders linear approximations prone to physically infeasible dispatch.

One approach to enforcing feasibility under nonconvex constraints is iterative projection using gradient information ~\cite{huang2024unsupervised}; however, nonconvexity precludes strict feasibility guarantees. Liang et al.~\cite{liang2024homeomorphic} propose a homeomorphic transformation enabling efficient bisection-based projection onto the feasible set; however, their method is restricted to simply connected sets, whereas the power flow feasible set is generally non-simply connected and often contains holes. To address this limitation, an interior-point (IP) NN is proposed to identify an IP of the nonconvex feasible set, upon which bisection-based projection is performed~\cite{liangefficient}. Nevertheless, the IP-NN lacks interpretability, and the identified IP cannot be guaranteed to lie strictly within the feasible region.

To the best of the authors’ knowledge, there remains a critical gap: \emph{\textbf{a solver-free neural dispatch framework that provides rigorous feasibility guarantees under nonconvex power flow constraints}}. To bridge this gap, this letter proposes a neural dispatch method for distributed energy resources with feasibility guarantees. A linear inner approximation of the DistFlow model is first derived via the convex envelopes theorem. Building on this approximation, a robust optimization-based affine policy is formulated to yield an IP mapping rule with strict theoretical guarantees. This rule is then embedded within a bisection-based projection that efficiently maps infeasible NN outputs onto the feasible set without external solvers.

\section{Problem Formulation}
Consider a radial balanced distribution network with $n+1$ buses indexed by $\mathcal{N}$, where node $0$ is the substation with fixed voltage magnitude $V_0$. Let $\mathcal{N}^+ = \mathcal{N} \setminus \{0\}$. The network is modeled via the DistFlow model, where $P_{ij}$ and $Q_{ij}$ denote branch power flows, $P_i$ and $Q_i$ denote the nodal injections, $V_i$ denotes the squared voltage, and $l_{ij}$ denotes the squared current. Branch has impedance $z_{ij} = r_{ij} + j x_{ij}$. The operator minimizes line losses and PV curtailment:
\begin{subequations}
\label{eq:distflow_model}
\begin{alignat}{2}
& \min_{\lambda_i, Q_i^{PV}} \sum_{ij \in \mathcal{L}} r_{ij} l_{ij} + \sum_{i \in \mathcal{N}} \left( \bar{P}_i^{PV} - P_i^{PV} \right) \label{eq:obj_vvo} \\
& \text{s.t.} ~~P_{ij} = P_j + \sum_{k:(j,k)\in \mathcal{L}} P_{jk} + r_{ij} l_{ij}, 
~ \forall (i,j) \in \mathcal{L} \label{eq:p_flow} \\
&  Q_{ij} = Q_j + \sum_{k:(j,k)\in \mathcal{L}} Q_{jk} + x_{ij} l_{ij}, 
~ \forall (i,j) \in \mathcal{L} \label{eq:q_flow} \\
&  V_i = V_j + 2(r_{ij} P_{ij} + x_{ij} Q_{ij}) - |z_{ij}|^2 l_{ij},~ \forall (i,j) \in \mathcal{L} \label{eq:voltage_drop} \\
&  l_{ij} V_i = P_{ij}^2 + Q_{ij}^2, ~ \forall (i,j) \in \mathcal{L} \label{eq:current_voltage} \\
&  \underline{V} \leq V_i \leq \overline{V},\forall i \in \mathcal{N}, ~~ l_{ij} \leq \bar{l}_{ij}, ~\forall (i,j) \in \mathcal{L} \label{eq:limit} \\
&  P_i = P_i^L - P_i^{PV},~Q_i =  Q_i^L - Q_i^{PV},~\forall i \in \mathcal{N}^+\label{eq:p_balance} \\
&  P_i^{PV} = \lambda_i \bar{P}_i^{PV}, 0 \leq \lambda \leq 1, ~\forall i \in \mathcal{N} ^+\label{eq:PV_P_limit} \\ 
&  \left(P_i^{PV}\right)^2 + \left(Q_i^{PV}\right)^2 \leq S_i, ~\forall i \in \mathcal{N}^+ \label{eq:inverter_limit}
\end{alignat}
\end{subequations}
where $\mathcal{L}$ is the set of branches, $\bar{P}_i^{PV}$ is the available PV generation, $\lambda$ is the PV curtailment ratio, $P_i^{PV}/Q_i^{PV}$ and $P_i^{L}/Q_i^{L}$ denote the active/reactive power outputs of the PV units and loads, respectively, $\bar{S}_i$ is the inverter capacity, $\bar{l}_{ij}$ is the branch current limit, and $\overline{V}$ and $\underline{V}$ are the voltage limits.

\section{Methodology}
% This section first presents the IP-based feasibility projection mechanism, followed by the proposed IP affine rule with strict theoretical guarantees.
\subsection{Interior-Point-Based Feasibility Projection}
The IP-based projection maps an infeasible NN output $f_{\mathrm{NN}}$ onto the feasible set using an interior point $f_{\mathrm{IP}}$ as follows:
\begin{align}
    \label{eq:proj}
\tilde{f}_{NN} = \kappa({f}_{NN} - {f}_{IP}) + {f}_{IP}
\end{align}
where $\kappa \in [0, 1]$ is the projection coefficient.

To determine a valid $\kappa$, we employ a bisection procedure that constructs a line segment between the NN output $f_{NN}$ and a strictly IP $f_{IP}$, iteratively refining the interval $[\kappa_l, \kappa_u]$ initialized to $[0, 1]$. At each iteration, the midpoint $\kappa = (\kappa_l + \kappa_u)/2$ is evaluated: if the interpolated point $\kappa(f_{NN} - f_{IP}) + f_{IP}$ is feasible, the lower bound is updated as $\kappa_l \gets \kappa$; otherwise, the upper bound is reduced as $\kappa_u \gets \kappa$. Upon convergence, $\kappa_l$ yields the feasible projection. If $f_{\mathrm{IP}}$ could be computed analytically at a cost comparable to NN inference, the projection would be highly efficient \cite{liang2024homeomorphic,liangefficient}. However, under nonconvex DistFlow constraints, such an analytical IP is not available. To address this challenge, we propose an inner convex approximation of the DistFlow model that yields a tractable affine IP with strict theoretical guarantees.

\subsection{Affine Interior-Point Rule via Inner Convex Approximation}
% To derive the inner convex approximation, \eqref{eq:p_flow}--\eqref{eq:voltage_drop} are cast in compact matrix form. Let $B$ denote the incidence matrix with entry $(i,j)$ equal to $1$ if node $i$ is connected to branch $j$ and $0$ otherwise. Define $C=(I_n-A)^{-1}$ with $A=\begin{bmatrix} 0_n & 1_n \end{bmatrix} B-I_n$, $D_R=CAR$, and $D_X=CAX$. This yields the affine voltage mapping:
% \begin{equation}
% V = V_0\mathbf{1}_n + M_p p + M_q q - Hl
% \label{eq:linear_voltage_model}
% \end{equation}
% where $M_p=2C^\top RC$, $M_q=2C^\top XC$, and $H=C^\top(2RD_R+2XD_X+Z^2)$.

% Let $l^b$ and $l^u$ denote the lower and upper bounds of $l$. The auxiliary variables $P^\pm$, $Q^\pm$, and $V^\pm$ bounding power flows and voltages over $l\in[l^b,l^u]$ are defined as follows:
% \begin{subequations}
% \label{eq:auxiliary_bounds}
% \begin{align}
% &P^+ = C p - D_R l^b, ~ P^-= C p - D_R l^u \label{eq:p_bounds} \\
% &Q^+ = C q - D_X^+ l^b - D_X^- l^u \\
% &Q^- = C q - D_X^+ l^u - D_X^- l^b \label{eq:q_UPbounds} \\
% &V^+ = V_0 \mathbf{1}_n + M_p p + M_q q - H^+ l^b - H^- l^u \label{eq:v_plus} \\
% &V^- = V_0 \mathbf{1}_n + M_p p + M_q q - H^+ l^u - H^- l^b \label{eq:v_minus}
% \end{align}
% \end{subequations}
% where $D_X^+$ and $H^+$ collect the nonnegative elements of $D_X$ and $H$, while $D_X^-$ and $H^-$ collect their negative elements. Enforcing $l^u\leq\bar{l}$, $\underline{V}\leq V^-$, and $V^+\leq\overline{V}$ guarantees satisfaction of the original limits.

Let $B$ denote the incidence matrix with entry $(i,j)$ equal to $1$ if node $i$ is connected to branch $j$ and $0$ otherwise. Define $A=\begin{bmatrix} 0_n & 1_n \end{bmatrix}B-I_n$, $C=(I_n-A)^{-1}$, $D_R=CAR$, and $D_X=CAX$, where $R=\mathrm{diag}\{r_{ij}\}$, $X=\mathrm{diag}\{x_{ij}\}$, and $Z^2=\mathrm{diag}\{|z_{ij}|^2\}$ collect the branch resistances and reactances. Let $P=[P_{ij}]$, $Q=[Q_{ij}]$, $p=[P_i]$, $q=[Q_i]$, $l=[l_{ij}]$, and $V=[V_i]$ denote the corresponding stacked vectors. Then \eqref{eq:p_flow}--\eqref{eq:voltage_drop} admit the compact matrix form \cite{9580648}:
\begin{subequations}
\label{eq:matrix_compact}
\begin{align}
P &= Cp - D_R l, \quad Q = Cq - D_X l \label{eq:PQ_compact}\\
V &= V_0\mathbf{1}_n + M_p p + M_q q - Hl \label{eq:V_compact}
\end{align}
\end{subequations}
where $M_p=2C^\top RC$, $M_q=2C^\top XC$, and $H=C^\top(2RD_R+2XD_X+Z^2)$.

Let $l^b$ and $l^u$ denote the lower and upper bounds of $l$. The auxiliary variables $P^\pm$, $Q^\pm$, and $V^\pm$ bounding power flows and voltages over $l\in[l^b,l^u]$ are defined as follows:
\begin{subequations}
\label{eq:auxiliary_bounds}
\begin{align}
&P^+ = C p - D_R l^b, ~ P^-= C p - D_R l^u \label{eq:p_bounds} \\
&Q^+ = C q - D_X^+ l^b - D_X^- l^u \\
&Q^- = C q - D_X^+ l^u - D_X^- l^b \label{eq:q_UPbounds} \\
&V^+ = V_0 \mathbf{1}_n + M_p p + M_q q - H^+ l^b - H^- l^u \label{eq:v_plus} \\
&V^- = V_0 \mathbf{1}_n + M_p p + M_q q - H^+ l^u - H^- l^b \label{eq:v_minus}
\end{align}
\end{subequations}
where $D_X^+$ and $H^+$ collect the nonnegative elements of $D_X$ and $H$, while $D_X^-$ and $H^-$ collect their negative elements. Enforcing $l^u\leq\bar{l}$, $\underline{V}\leq V^-$, and $V^+\leq\overline{V}$ guarantees satisfaction of the original limits.

Because the Hessian of \eqref{eq:current_voltage} is positive semi-definite, its second-order expansion is convex and the first-order Taylor expansion provides a global underestimator \cite{9580648}:
\begin{equation}
l_{ij}^{b} \leq l_{ij}^0 + J_{ij+}^\top \delta_{ij}^+ + J_{ij-}^\top \delta_{ij}^- \leq l_{ij}
\label{eq:lower_bound_l}
\end{equation}
where $J_{ij+}$ and $J_{ij-}$ contain the positive and negative elements of Jacobian matrices of \eqref{eq:current_voltage}, $\delta_{ij}^+ = (P_{ij}^+, Q_{ij}^+, V_i^+)$ and $\delta_{ij}^- = (P_{ij}^-, Q_{ij}^-, V_i^-)$.

The upper bound follows from epigraph relaxation as:
\begin{equation}
l \leq \big((P^*)^2+(Q^*)^2\big)/\underline{V} \leq l^u
\label{eq:soc_relaxation}
\end{equation}
with $(P^*,Q^*)$ taking four combinations: $(P_{ij}^+, Q_{ij}^+)$, $(P_{ij}^+, Q_{ij}^-)$, $(P_{ij}^-, Q_{ij}^-)$, and $(P_{ij}^-, Q_{ij}^+)$.

Consequently, \eqref{eq:p_flow}--\eqref{eq:limit} reduce to \eqref{eq:auxiliary_bounds}--\eqref{eq:soc_relaxation} subject to $l^u\leq\bar{l}$, $\underline{V}\leq V^-$, and $V^+\leq\overline{V}$, yielding a convex inner approximation with decision variables comprising the dispatched PV active/reactive power and current envelopes $l^u$, $l^b$.

Let $x$ denote the vector of active/reactive loads and maximum PV active power, taking values in $\mathcal{X} := \{x \mid \underline{x} \leq x \leq \overline{x}\}$. We adopt an affine IP rule with respect to $x$:
\begin{equation}
\tilde{f}_{\mathrm{IP}}(x) = W_{\mathrm{IP}} x + w_{\mathrm{IP}}
\label{eq:affine_rule}
\end{equation}
where $W_{\mathrm{IP}}$ and $w_{\mathrm{IP}}$ denote the affine coefficients, and $\tilde{f}_{\mathrm{IP}}(x) = [f_{\mathrm{IP}}(x)^\top, (l^b)^\top, (l^u)^\top]^\top$ augments the IP with current envelope variables. During online operation, only the coefficients for $f_{\mathrm{IP}}(x)$ are extracted to compute the IP.

The affine coefficients $\gamma = (W_{\mathrm{IP}}, w_{\mathrm{IP}})$ must be chosen such that the resulting IP satisfies all constraints for every $x \in \mathcal{X}$. This requirement embodies the core principle of robust optimization, motivating the formulation of the following compact robust optimization problem for determining $\gamma$:
\begin{subequations}
\label{eq:optimization_problem}
\begin{align}
&\max_{\gamma, s \geq 0} \; s \label{eq:obj} \\
&\text{s.t.} ~ A_l(\gamma)x - b_l(\gamma) + s \leq 0, ~ \forall x \in \mathcal{X} \label{eq:lc} \\ 
& ~~~~\|A_s(\gamma) x + b_s(\gamma)\|_2 + s \leq c_s(\gamma), ~ \forall x \in \mathcal{X} \label{eq:soc}
\end{align}
\end{subequations}
where \eqref{eq:lc} denotes the linear constraints, \eqref{eq:soc} represents the second-order cone (SOC) constraints, and the slack variable $s$ quantifies the distance to the constraint boundaries.

The linear constraints \eqref{eq:lc} are reformulated via duality as
\begin{subequations}
\label{eq:linear_duality}
\begin{align}
&\bar{x}^{\top}\lambda_{l}^{+} - \underline{x}^{\top}\lambda_{l}^{-} + s \leq b_{l}(\gamma)\\
&\lambda_{l}^{+} - \lambda_{l}^{-} = A_{l}(\gamma)^{\top}
\end{align}
\end{subequations}
where $\lambda_{l}^{+} \geq 0$ and $\lambda_{l}^{-} \geq 0$ are the dual variables associated with $x \leq \bar{x}$ and $\underline{x} \leq x$, respectively.

For the SOC constraints \eqref{eq:soc}, the norm inequality $\|\cdot\|_2 \leq \|\cdot\|_1$ yields the approximation:
\begin{equation}
\|A_{s}(\gamma)x + b_{s}(\gamma)\|_{1} + s \leq c_{s}(\gamma), ~ \forall x \in \mathcal{X}.
\label{eq:soc_conservative}
\end{equation}

Strong duality converts \eqref{eq:soc_conservative} to a tractable linear model, paralleling the treatment of \eqref{eq:lc} in \eqref{eq:linear_duality}. Solving the tractable reformulation of \eqref{eq:optimization_problem} yields affine decision rules enabling rapid IP generation with strict feasibility guarantees.

\subsection{Neural Network Training}
To align with the IP formulation, the NN is designed to output solely the dispatch decisions, namely the PV active and reactive power setpoints. Consequently, the system state variables $z$, comprising voltages and branch currents, are recovered implicitly via the power flow equations $h_{\mathrm{pf}}(f_{\mathrm{NN}}, z)=0$. By the implicit function theorem, the state sensitivity is
\begin{equation}
\frac{dz}{df_{\mathrm{NN}}} = -\left(\frac{\partial h_{\mathrm{pf}}}{\partial z}\right)^{-1}\frac{\partial h_{\mathrm{pf}}}{\partial f_{\mathrm{NN}}}.
\label{eq:dz_dfn}
\end{equation}

Denote the NN output as $f_{\mathrm{NN}}(x;\theta)$, taking $x$ as the input vector with parameters $\theta$. The training loss combines supervised learning against optimal labels $f^*$ with penalties on constraint violations evaluated at the recovered state variables:
\begin{multline}
\mathcal{L} = \|f_{\mathrm{NN}}(x;\theta) - f^*\|_2^2 + \pi_v \sum_{i}\max(0, \underline{V} - V_i) 
\\ + \pi_v \sum_{i}\max(0, V_i - \overline{V}) + \pi_l \sum_{ij}\max(0, l_{ij} - \bar{l}_{ij})
\label{eq:loss}
\end{multline}
where $\pi_v$ and $\pi_l$ penalize system state constraint violations.

The gradient of \eqref{eq:loss} with respect to  parameters $\theta$ follows from the chain rule:
\begin{equation}
\nabla_{\theta}\mathcal{L} = \frac{\partial \mathcal{L}}{\partial f_{\mathrm{NN}}}\frac{df_{\mathrm{NN}}}{d\theta} + \frac{\partial \mathcal{L}}{\partial z}\frac{dz}{df_{\mathrm{NN}}}\frac{df_{\mathrm{NN}}}{d\theta}
\label{eq:gradient}
\end{equation}
where $\frac{dz}{df_{\mathrm{NN}}}$ is given by \eqref{eq:dz_dfn} and all remaining terms are analytically tractable via automatic differentiation. Supervised pre-training without penalty terms is adopted to prevent power flow infeasibility caused by poor initialization.

\section{Case Study}
We evaluate the proposed method on 33-bus and 129-bus networks with 7 and 28 PV units. Comparisons are conducted against both soft-penalty-based and theoretically guaranteed benchmarks: (i) V-NN, a vanilla NN trained via supervised learning on IPOPT-generated optimal solutions; (ii) P-NN, augmenting V-NN with penalty-based constraint handling per \cite{huang2024unsupervised}; (iii) O-NN, projecting infeasible P-NN outputs via orthogonal projection using IPOPT; and (iv) B-NN, the proposed approach. Performance is evaluated in terms of optimality, feasibility, and computational efficiency. A total of 7,000 samples are generated by perturbing nominal parameters within $\pm$25\% and partitioned into training, validation, and test sets in a 5:1:1 ratio. All methods employ fully connected networks with two hidden layers, and their hyperparameters are carefully tuned. All data and source codes are available at \url{https://github.com/JieZhu6/RTN-DER-DFG}.

Table~\ref{tab:33_bus_comparison} presents the comparative results on the 33-bus system. The proposed B-NN overcomes the feasibility limitations of soft-penalty-based P-NN through its IP-based projection. Unlike O-NN, which requires solving an external optimization problem for orthogonal projection, the proposed method employs a analytical projection that incurs only 0.059\% optimality loss while achieving a 28$\times$ speedup, yielding projection times comparable to NN inference. Table~\ref{tab:129_bus_comparison} presents results on the larger-scale 129-bus system. The solver-free B-NN achieves a substantial projection-time speedup over O-NN with limited optimality loss.

\begin{table}
\caption{Comparison of Different Methods in 33-bus Network}
\label{tab:33_bus_comparison}
\centering
{
\begin{tabular}{ccccc}
\hline
Method & \makecell{Optimal\\gap (\%)} & \makecell{Feasibility\\rate (\%)} & \makecell{Inference\\Time (s)} &\makecell{Projection\\Time (s)} 
\\
\hline
V-NN & 0.082 & 47.70 &  0.0011  & --  
\\
P-NN& 1.711 & 87.20 &  0.0011  & -- 
\\
O-NN& 1.890 & 100.00 & 0.0011  & 0.0894  
\\
B-NN& 1.949 & 100.00 &  0.0011  & 0.0032  \\
\hline
\end{tabular}
}
\end{table}

\begin{figure}[!t]
   \centering
   \includegraphics[width=0.45\textwidth]{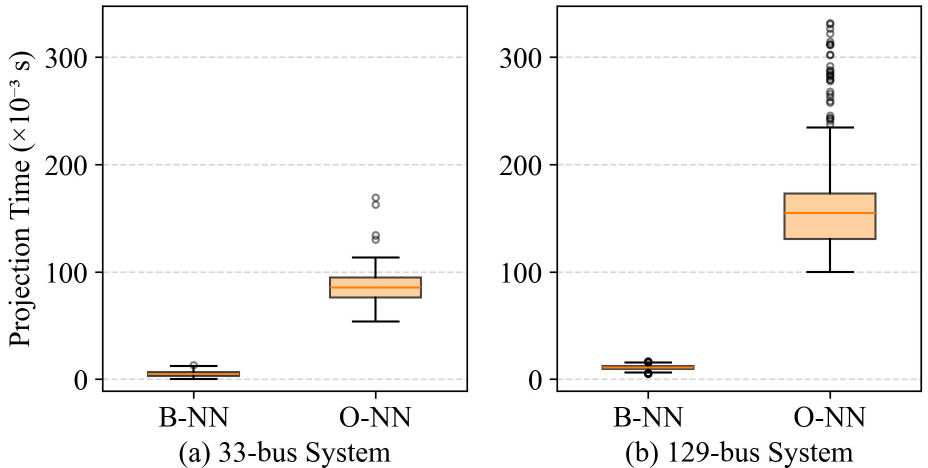}
   \caption{Boxplot of projection times across 1,000 test samples.}
   \label{fig:Boxplot}
\end{figure}

Fig.~\ref{fig:Boxplot} compares projection time distributions across infeasible samples: B-NN maintains consistently low and stable times, whereas O-NN exhibits high variance due to its reliance on external solvers for nonconvex problems. This confirms that the proposed method achieves both reliability and efficiency.

\begin{table}
\caption{Comparison of Different Methods in 129-bus Network}
\label{tab:129_bus_comparison}
\centering
{
\begin{tabular}{ccccc}
\hline
Method & \makecell{Optimal\\gap (\%)} & \makecell{Feasibility\\rate (\%)} & \makecell{Inference\\Time (s)} &\makecell{Projection\\Time (s)} 
\\
\hline
V-NN & 0.429 & 8.81 &  0.0012  & --  
\\
P-NN& 3.189 & 69.37 &  0.0012  & -- 
\\
O-NN& 3.373 & 100.00 & 0.0012  & 0.1588  
\\
B-NN& 3.717 & 100.00 &  0.0012  & 0.0071  \\
\hline
\end{tabular}
}
\end{table}

\section{Conclusion}
This letter proposes a solver-free neural dispatch framework that enforces strict feasibility under nonconvex power flow constraints. By integrating a robust affine interior-point rule with bisection-based projection, the method achieves feasibility recovery on par with NN inference while maintaining near-optimal performance, effectively bridging real-time dispatch and rigorous physical feasibility.

\bibliographystyle{IEEEtran}
\bibliography{ref}

\end{document}